\documentclass[10pt,letterpaper]{article}
\usepackage{opex3}
\usepackage{color}
\usepackage{amsmath,cite}

\begin{document}

\title{Experimental generation of 8.4 dB entangled state with an optical cavity involving a wedged type-II nonlinear crystal}

\author{Yaoyao Zhou, $^1$ Xiaojun Jia,$^{1,*}$ Fang Li, $^1$ Changde Xie,$^{1}$ and Kunchi Peng$^{1}$}

\address{$^1$State Kay Laboratory of Quantum Optics and Quantum Optics Devices,\\ Institute of Opto-Electronics, Shanxi University, Taiyuan, Shanxi, 030006, China}

\email{$^*$jiaxj@sxu.edu.cn} 



\begin{abstract}
Entangled state of light is one of the essential quantum resources
in quantum information science and technology. Especially, when the fundamental principle experiments have been achieved in labs and the applications of continuous variable quantum information in the real world are considered, it is crucial to design and construct the generation devices of entangled states with high entanglement and compact configuration. We have designed and built an efficient and compact light source of entangled state, which is a non-degenerate optical parametric amplifier (NOPA) with the triple resonance of the pump and two subharmonic modes. A wedged type-II KTP crystal inside the NOPA is used for implementing frequency-down-conversion of the pump field to generate the optical entangled state and achieving the dispersion compensation between the pump and the subharmonic waves. The EPR entangled state of light
with quantum correlations of 8.4 dB for both amplitude and phase
quadratures are experimentally produced by a single NOPA under the
pump power of 75 mW.
\end{abstract}

\ocis{(270.0270) Quantum optics; (270.6570) Squeezed states;
(190.4970) Parametric oscillators and amplifiers.} 


\section{Introduction}

After continually exploring in the past few decades, the great
potentiality and scientific significance of quantum information have
been extensively recognized. Utilizing discrete variables (dv) of single photons
(qubits) and continuous variables (cv) of optical fields (qumodes)
to be the information carriers, two types of optical quantum
information technology have been parallelly developing. A variety of
novel features and advantages of quantum communication and quantum
computation over classical physics have been exhibited in both dv
and cv regimes\cite {DBou,DBosch,LVaidman,SL,AFuru}.

Entangled state of light with quantum correlations of amplitude and
phase quadratures is the necessary quantum resources in optical cv
quantum information systems. Especially, when the fundamental principle experiments have been achieved in labs \cite{Yuka08,AFuru,Jia} and the applications of cv quantum information in the real world are considered, it is crucial to design and construct the generation devices of entangled states with high entanglement and compact configuration. Degenerate and non-degenerate optical parametric amplifiers (DOPAs
and NOPAs) consisting of type-I and type-II nonlinear crystals,
respectively, are two kinds of optical devices for successfully
generating cv squeezed and entangled states of optical
field\cite{Ou,Mehmet,Morin2,TEberle,WP,JMizuno,zhangwang,JLaurat,Nobuyuki}.
Two subharmonic modes (signal and idler) produced by a DOPA are
totally degenerate which form a single-mode squeezed state.
Interfering two single-mode squeezed states generated by two DOPAs
with identical configuration on a $50/50$ beamsplitter, the
two output optical modes from the beamsplitter construct a cv
entangled state\cite{TEberle,WP,JMizuno}. Because the nonlinear
coefficient of type-I crystal is higher than that of type-II crystal
generally, the single-mode squeezed state over 10 dB has
been generated by DOPA\cite{Mehmet,Morin2}. The entangled state of
about 10 dB has been also obtained with a pair of DOPAs and a
beamsplitter\cite{TEberle}. On the other hand, the signal and idler
modes produced by a NOPA are nondegenerate with perpendicular
polarizations and construct a cv entangled state directly\cite {Ou}. Comparing with the entanglement generation system of interfering two single-mode squeezed states produced by two separated DOPAs on a 50/50 beam splitter, the single NOPA is quite compact.
Although Einstein-Podolsky-Rosen (EPR) entangled state can be
prepared by a single NOPA, the obtained entanglement degree kept
lower than 4 dB for a long time since it was generated firstly by
Kimble's group at the beginning of 1990s due to some technic
troubles\cite{zhangwang,JLaurat,Nobuyuki}. Until 2010, our group
improved the entanglement degree to 6 dB by decreasing the extra
noise of the pump laser and reducing the phase fluctuations in phase
locking systems\cite{WangShen}. Then the EPR entanglement of 8.1 dB
was demonstrated by a cascaded system consisting of three
NOPAs\cite{zhuhui}. Very recently, the hybrid entanglement between
particale-like and wave-like optical qubits has been experimentally
achieved by combining a type-I and type-II optical parametric
oscillators\cite{Morin}. Here, we present a new experimental result
in which the cv entanglement of 8.4 dB is directly produced by a
simple and compact triple-resonant NOPA with a wedged type-II KTP crystal under
75 mW pump power.

For a NOPA operating below its oscillation threshold and in the
state of deamplification (the phase difference between the pump
field and the sum of the signal phase and the idler phase equals to $(2n+1)\pi $, $n$ is an
integer), the correlation
variances ($\langle \delta ^{2}(\hat{X}_{a_{_{1}}}^{out}+\hat{X}%
_{a_{_{2}}}^{out})\rangle $, $\langle \delta ^{2}(\hat{Y}_{a_{_{1}}}^{out}-%
\hat{Y}_{a_{_{2}}}^{out})\rangle $) and anti-correlation variances
($\langle
\delta ^{2}(\hat{X}_{a_{_{1}}}^{out}-\hat{X}_{a_{_{2}}}^{out})\rangle $, $%
\langle \delta ^{2}(\hat{Y}_{a_{_{1}}}^{out}+\hat{Y}_{a_{_{2}}}^{out})%
\rangle $) of amplitude and phase quadratures between the output
signal and idler modes in ideal condition are expressed
by\cite{Mehmet,zhuhui}:

\begin{eqnarray}
&&\langle \delta ^{2}(\hat{X}_{a_{_{1}}}^{out}\pm \hat{X}_{a_{_{2}}}^{out})%
\rangle =\langle \delta ^{2}(\hat{Y}_{a_{_{1}}}^{out}\mp \hat{Y}%
_{a_{_{2}}}^{out})\rangle  \\
&=&2(1\mp \eta _{\det }\eta _{esc}\frac{4\sqrt{P/P_{thr}}}{(1\pm \sqrt{%
P/P_{thr}})^{2}+4(2\pi f\kappa ^{-1})^{2}}),  \notag
\end{eqnarray}
where, $\hat{X}_{a_{_{1(2)}}}^{out}$ and $\hat{Y}_{a_{_{1(2)}}}^{out}$ are
the amplitude and the phase quadratures of the output signal (idler) mode
from the NOPA, respectively; $\eta _{\det }$ is the imperfect detection
efficiency ($\eta _{\det }\leq 1$); $\eta _{esc}=T/(T+L)$ is the escape
efficiency, $T$ is the transmissivity of the output coupler of the NOPA for
the subharmonic modes and $L$ stands for all extra intracavity losses; $f$
and $\kappa $ are the noise analysis frequency and the decay rate of the
NOPA cavity, respectively; $P$ is the pump power and $P_{thr}$
is the oscillation threshold of the NOPA, which can be estimated from the formula%
\cite{Fabre}:

\bigskip
\begin{equation}
P_{thr}=\frac{(T_{0}+L_{0})^{2}(T+L)^{2}}{8\chi ^{2}T_{0}}
\end{equation}
where $T_{0}$ and $L_{0}$ stand for the transmissivity of the input coupler of the
NOPA and total intracavity losses for the harmonic mode, respectively. $\chi $
is the nonlinear coefficient of the crystal.

When the second term in Eq. (1) equals zero, there is no quantum correlation
between the output signal and idler modes and the correlation variance
equals "2", which is the normalized shot noise limit (SNL) of the
correlation variance. If the second term is between $0$ and $1$, the
correlation variance ($\langle \delta ^{2}(\hat{X}_{a_{_{1}}}^{out}+\hat{X}%
_{a_{_{2}}}^{out})\rangle $, $\langle \delta ^{2}(\hat{Y}_{a_{_{1}}}^{out}-%
\hat{Y}_{a_{_{2}}}^{out})\rangle $) is smaller than the corresponding SNL.
In this case the output field is a cv entangled state of the optical field. The larger the
second term is, the higher the entanglement degree is. From Eq. (1) we can see
that increasing $\eta _{esc}$, i.e. increasing $T$ and decreasing $L$, is
the most efficient way to obtain high entanglement. However, Eq. (2) shows that the threshold power $P_{thr}$ of the NOPA depends on not
only the transmissivity of the output coupler of the NOPA for the subharmonic
modes ($T$) but also the transmissivity of the input coupler for
the harmonic modes ($T_{0}$). Increasing $T$ the threshold $%
P_{thr}$ must increase. To maintain a lower $P_{thr}$, $T_{0}$ has to be decreased. In
this case the operating model of triple-resonance of signal, idler and pump
modes should be a favorable choice. For completing
three-mode resonance, two physical parameters of NOPA need to be adjusted at
least\cite{Guo,Juwi,Zhangmin,Auria,Gross,Stefszky,Ime,Chua}. By adjusting the temperature of KTP around the phase-match
point, the double-resonance of the signal and idler modes can be demonstrated
\cite{Ou,WangShen}. In \cite{Guo,Gross,Stefszky}, an extra optical element is inserted in the optical cavity to compensate the dispersion between the pump and the subharmonic modes. However, because the inserted element must increase the intracavity loss of the NOPA, the high entanglement is not obtained \cite{Guo}. An other scheme of the dispersion compensation is to transversally move a wedged crystal in the optical cavity and realize the compensation of the optical paths between different wavelength \cite{Auria,Ime,Chua}. To compensate the dispersion and not increase the loss, we cut a
wedge in an end-face of the KTP crystal and move the crystal transversal to its y axis
in the NOPA to accomplish the dispersion compensation between the pump and
the subharmonic modes. In order to realize triple resonance simultaneously in the  NOPA
the optical paths $l(n_{j},d)$ traveled back and forth for the three optical modes should be an integer ($A_{j}$) multiple of their wavelength ($\lambda _{j}$):

\begin{equation}
l(n_{j},d)=2n_{j}\ast (l_{x}-d\ast \tan \theta )+2(l_{air}+d\ast
\tan \theta )=A_{j}\ast \lambda _{j}
\end{equation}%
where, $j=s$, $i$, $p$ ($s,$ $i,$ $p$ stands for signal, idler and pump
mode, respectively.); $n_{j}$ is
KTP refractive index for different modes \cite{BF,DW,Bier}; $l_{x}$ and $l_{air}
$ are the geometric length of the KTP and the distance traveled in the air in the NOPA
cavity, respectively. $d$ is the distance moved from one side of the crystal; $\theta $ is the wedged angel cut on the crystal.

For a Gaussian mode, the round-trip phase variation $\Phi $ undergone by a
cavity mode is given by \cite{Auria}

\begin{equation}
\Phi (n_{j},d)=\frac{2\pi }{\lambda _{j}}l(n_{j},d)-4\Phi
_{G}(n_{j})
\end{equation}
where $\Phi _{G}(n_{j})$ is the Gouy phase \cite{Yariv}. Triple
mode resonance occurs only when the conditions $%
\Phi (n_{j},d)=2\pi m_{j}$ with integer $m_{j}$ ($j=s,$ $i,$ $p$)
are simultaneously satisfied. In the experiment, the condition is met by precisely adjusting the temperature and transversally moving the positin of the KTP crystal.

\section{Experimental setup and results}

The schematic of the experimental setup is shown in Fig. 1. The laser
source was a continuous wave Nd: YAP/LBO laser with the
single-frequency outputs of both 1080 nm and 540 nm wavelengths,
which was provided by YuGuang company (CDPSSFG-VIB). The two output
laser were separated into two parts by a two color mirror with a
high-reflection (HR) coated for 540 nm and an anti-reflection (AR)
coated for 1080 nm. The light at the wavelength of 1080 nm was
transmitted through a mode cleaner (MC1) that implemented the
spatio-temporal filtering to improve the mode quality and the
stability of the injected signal of NOPA and the local oscillator (LO) beams of the
balanced homodyne detectors (BHD). The other MC (MC2) was used for
reducing extra amplitude and phase fluctuations of the pump field. MC1 and
MC2 were optical ring cavities consisting of three mirrors and their
finesses were 500 and 650, respectively. In the experiment the
lengths of MC1 and MC2 were servo-controlled to resonate with their
carrier field by a lock-in amplifier. For achieving locking of MC1 (MC2), we adhered a PZT6 (PZT7) on a mirror of MC and drove the PZT with electronic sine signal of about 9 kHz. The error signal used for the cavity locking is derived from the light leakeage of a mirror placed behind the MC. The cleaned laser at 540 nm was injected into the NOPA as the pump field. The seed beams including both horizontal polarization and the vertical polarization at 1080 nm were modulated by the phase electro-optic modulator (EOM) and were utilized to lock the NOPA cavity. The two subharmonic modes with the horizontal and vertical polarizations were adjusted to resonate in the NOPA by tuning the temperature of the nonlinear crystal KTP, firstly. Then the error signal for locking the length of the NOPA cavity was extracted through the detecter (D0), which detected the beam reflected by the isolator (ISO) placed in the optic path. At last, the error signal was fed back to PZT1 for locking the NOPA.

\begin{figure}[htb]
\centerline{\includegraphics[width=7.5cm]{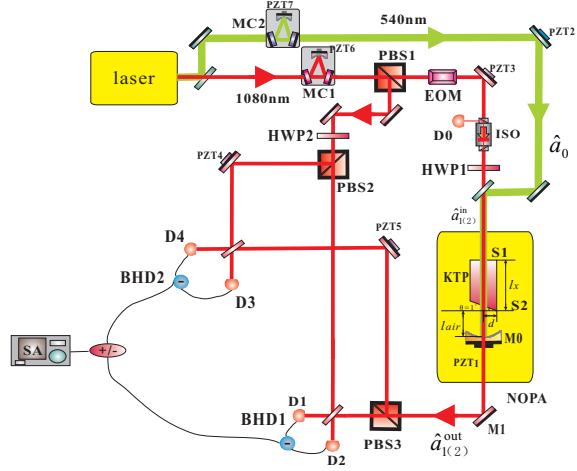}}
\caption{Experimental setup of the NOPA. Laser: Nd: YAP/LBO laser
source; MC1: mode cleaner of 1080 nm; MC2: mode cleaner of 540 nm;
HWP1-2: $\lambda/2 $ waveplate; PBS1-3: polarizing beam splitter; EOM: phase electro-optic modulator; ISO:
isolator; PZT1-7: piezoelectric transducer; D0-4: detector; BHD1-2:
balanced homodyne detector; SA: spectrum analyzer.}
\end{figure}

To avoid the walk-off of ordinary and extraordinary beams within the crystal, we choose the $\alpha$-cut KTP as nonlinear crystal in the experiment because it can realize type-II noncritical phase matching at 1080 nm \cite{Ou92}, that is why we choose the pump  laser of 540 nm and the injected beams of 1080 nm. The NOPA was a half-monolithic optical resonator built by a $\alpha$-cut KTP crystal of dimensions $3\times
3\times 10$ $mm^{3}$ and a
piezo-actuated output coupler (M0) which is a concave mirror with a
radius of curvature of 50 mm coated with $T=12.5\%$ for 1080 nm and
HR for 540 nm. The front face (S1) of the crystal was HR coated
for 1080 nm and $T_0=20\%$ coated for 540 nm, which served as the
input coupler. The end-face (S2) of the KTP was cut to $1^{\circ }$ along
y-z plane of the crystal and was AR coated for both 1080 nm and 540
nm. The crystal was placed in a copper-made oven, which can be
precisely temperature-controlled and slowly moved along its y axis. The
geometric length of the NOPA cavity was 54 mm. The measured finesse
of the NOPA ($49$) means that the total extra loss of the NOPA ($L$)
was $0.3\%$. At the case of triple-resonance the oscillation
threshold power of the NOPA was about 150 mW. With the pump power of
75 mW and the injected signal and idler modes of 10 mW, the two
subharmonic fields were generated through the intracavity frequency
down-conversion process. The process to realize the triple-resonance
of pump, signal and idler modes in the NOPA is exhibited in Fig. 2
when the length of the NOPA cavity is scanned. The upper traces in
Fig. 2 are the voltage signal on PZT1 for scanning the cavity length
of NOPA, the middle and lower traces are the transmissive modes of
the harmonic mode and subharmonic mode from the NOPA, respectively.
Figure 2(a) shows that the pump, signal and idler modes are not in the resonance (The three peaks appear at different time.). By tuning the temperature of the KTP crystal around 63 $^{\circ }C$, the double resonance of the signal and idler mode is achieved, as shown Fig. 2(b), where the two peaks for signal and idler modes overlap. We can see that the peak height in the lower trace of Fig. 2(b) is twice of that in Fig. 2(a), i.e. the peak intensity after achieving the double-resonance is twice of that of the individual peak before the double-resonance. Then we slowly moved the position of KTP crystal transversal to its y axis to meet the condition of the triple-resonance [Fig. 2(c)], the classical gain of the NOPA increases to 30 times quickly and the peak in the lower trace of Fig. 2(c) is also significantly increased.

\begin{figure}[htb]
\centerline{\includegraphics[width=7.5cm]{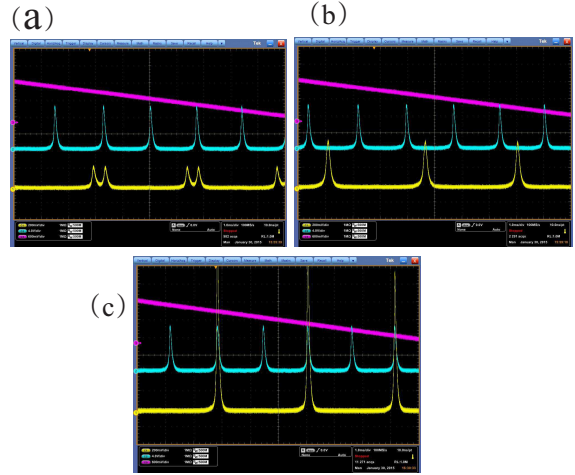}} \caption{The
process for the experimental realization of three modes resonance
in the NOPA when the length of NOPA is scanned. The uper trace for
the signals when the cavity length of the NOPA is scanning, the middle
trace for the pump modes, the lower trace for the signal and idler
modes.}
\end{figure}

\begin{figure}[htb]
\centerline{\includegraphics[width=7.5cm]{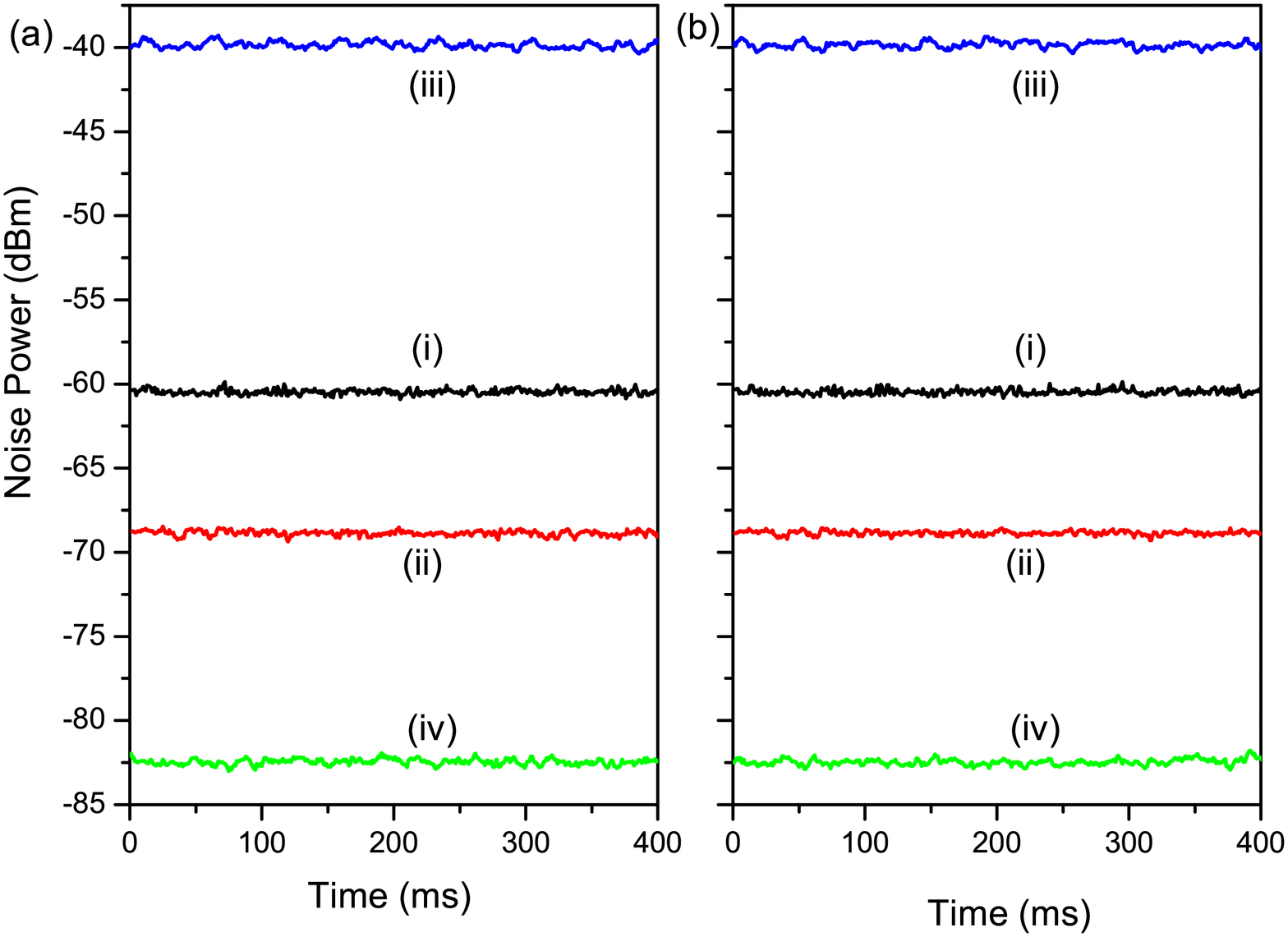}} \caption{The
measured correlation and anti-correlation variances noise powers of the EPR beams, where (i) is the SNL; (ii) is the
correlation variance of amplitude-sum (a) and phase-difference (b); (iii) is the
anti-correlation variance of amplitude-difference (a) and phase-sum (b); (iv) is the ENL. The measurement parameters of SA:\ RBW
10kHz; VBW 100Hz.}
\end{figure}

In the experiment, for directly measuring the entanglement between signal and idler modes, the two output modes are separated by PBS3 firstly and then their amplitude or phase quadratures are measured by two sets of BHD system at the same time. Each BHD system consists of a 50/50 beam splitter and two high quantum efficiency InGaAs photo detectors. When we measure the amplitude (phase) quadratures, the relative phases between the LO beam and the signal (idler) mode in both BHD1 and BHD2 should be locked at $n\pi $ ($(2n+1)\pi /2$) simultaneously. Then the outputs of two sets of BHD system are combined by the positive or negative power combiner to obtain the correlation (anti-correlation) variances of the sum or difference between the amplitude (phase) quadratures of the output signal and idler modes as needed. The correlation (anti-correlation) noises of the amplitude-sum (difference) and the phase-difference (sum)
were recorded by a spectrum analyzer (SA), respectively. The
measured noise powers of the quadrature amplitude (a) and the quadrature
phase (b) between the output signal and idler fields at 2MHz are shown in Fig. 3. Traces (i) are the corresponding
SNL, which were obtained by blocking the output beam of the NOPA.
The measured correlation
variances of the amplitude-sum and the phase-difference [as shown traces (ii) in Figs. 3(a) and 3(b)] are
$8.40\pm 0.18$ dB and $8.38\pm 0.16$ dB below the corresponding SNL,
respectively, which is in good agreement with that calculated by Eq. (1) when the non-perfect quantum efficiency of the BHD (95.0\%) and the escape efficiency (97.6\%) are considered. Traces (iii) in Figs. 3(a) and 3(b) are the measured anti-correlation
variances of the amplitude-difference and the phase-sum, which all are more than $20$ dB above the SNL. The traces (iv) are the
electronics noise level (ENL) of the photo detector, which were obtained by blocking the injected light field.

\section{Conclusion}

For the conclusion, we have designed and built an efficient NOPA and
experimentally obtained the cv optical entangled state of 8.4 dB. By means of a wedged nonlinear
type-II crystal the triple-resonance of the pump, signal and idler
modes in a NOPA is realized. Because there is no any additional
element in the optical cavity the intracavity loss is reduced to the
largest extent. Comparing the NOPA cavity with the previous one
containing an additional element in\cite{Guo}, its extra loss had
been reduced from 2.3\% to 0.3\%. According to the reduced loss, the
theoretically calculated correlation degree from Eq. (1) could be
improved from 2.4 dB to 9.0 dB in the ideal condition. However, in
the real experiment, due to the imperfectly relative phase-locking
between the pump field and injected signal as well as the local beam
and the measured signal on each BHD, the large anti-correlation
variances will inevitably effect the measured results of correlation
variance, and thus the measured correlation level is lower than the
ideal values\cite{Mehmet}. The triple-resonance scheme significantly
decreases the oscillation threshold of NOPA than that of
double-resonance. The oscillating threshold of the NOPA is 150 mW in
the case of the three-mode resonance, which is in reasonable agreement
with the calculation with Eq. (2). For the
double-resonance NOPA, the threshold will increase to about 1875 mW,
which is estimated from Eq. (2). The
low threshold is specially useful for cv quantum networks in which
multipartite entangled states are required\cite{Su,Aoki,Su12,Tan}.
Cv multipartite entangled states are generated by the interferences
of squeezed states more than one on a beam-splitter network
\cite{Loock00}. The four-partite entangled states have been prepared
by interfering four squeezed states, which were produced by four
DOPAs\cite{MYu} or two NOPAs\cite{Su}, respectively. In these
systems all DOPAs or NOPAs were pumped by a laser to keep the
classical coherence among the interfered squeezed states. Therefore
it is important to reduce the pump threshold of each parametric
amplifier when we use an available laser power to generate
multipartite entangled states with more submodes. The presented
experiment shows that high cv entanglement is scalable with a single
NOPA.

\section*{Acknowledgments}

This research was supported by Natural Science Foundation of China
(Grants Nos. 11322440, 11474190, 11304190, 61121064), FOK YING TUNG Education
Foundation, Shanxi Province Science Foundation(Grants No. 2014021001) and National Basic Research Program of China (Grant No.
2010CB923103).

\end{document}